\documentclass[aps,prl,twocolumn,10pt,superscriptaddress]{revtex4-1}
\usepackage[pdftex]{color,graphicx}

\begin{document}

\title{Structural, mechanical, and electronic properties of single graphyne layers based on a 2D biphenylene network}

\author{Mateus Silva Rêgo}
\affiliation{Departamento de F\'isica, Universidade Federal do Piau\'i, CEP 64049-550, Teresina, Piau\'i, Brazil}
\author{M\'ario Rocha dos Santos}
\affiliation{Departamento de F\'isica, Universidade Federal do Piau\'i, CEP 64049-550, Teresina, Piau\'i, Brazil}
\author{Marcelo Lopes Pereira Júnior}
\affiliation{University of Brasília, College of Technology, Department of Electrical Engineering, Brasília, Brazil} 
\author{Eduardo Costa Gir\~ao}
\affiliation{Departamento de F\'isica, Universidade Federal do Piau\'i, CEP 64049-550, Teresina, Piau\'i, Brazil} 
\author{Vincent Meunier}
\affiliation{Department of Engineering Science and Mechanics, Pennsylvania State University, State College, PA, USA}
\author{Paloma Vieira Silva}
\affiliation{Departamento de F\'isica, Universidade Federal do Piau\'i, CEP 64049-550, Teresina, Piau\'i, Brazil}

 \date{\today}
\begin{abstract}
Graphene is a promising material for the development of applications in nanoelectronic devices, but the lack of a band gap necessitates the search for ways to tune its electronic properties. In addition to doping, defects, and nanoribbons, a more radical alternative is the development of 2D forms with structures that are in clear departure from the honeycomb lattice, such as graphynes, with the distinctive property of involving carbon atoms with both hybridizations $sp$ and $sp^2$. The density and details of how the acetylenic links are distributed allow for a variety of electronic signatures. Here we propose a graphyne system based on the recently synthesized biphenylene monolayer. We demonstrate that this system features highly localized states with a spin-polarized semiconducting configuration. We study its stability and show that the system's structural details directly influence its highly anisotropic electronic properties. Finally, we show that the symmetry of the frontier states can be further tuned by modulating the size of the acetylenic chains forming the system.
\end{abstract}

\maketitle

\section{Introduction}
The nanoscale systems studied in the literature comprise a large number of structures with several different chemical compositions and structural arrangements~\cite{song2010,lin2013,bianco2013,geim2013,das2014,algara2014,zhu2015}. Carbon-based nanostructures are key examples of such materials for historical and scientific reasons. On the one hand, the development timeline of this research field is closely related to nanocarbon forms such as fullerenes, nanotubes, and graphene~\cite{noorden2011}. On the other hand, carbon nanostructures have physical and chemical properties that can be modulated over a wide range of possibilities due to the details of their atomic structures~\cite{dai2002,jia2011,terrones2012,meunier2016}.

Today, graphene is the most iconic representative of the nanocarbon family. However, many forms based on chemical and/or physical modifications of graphene have been studied as motivated by the need to open an electronic band gap, absent in pristine graphene. These include, but are not limited to, doping~\cite{cruzsilva2011,lee2018}, insertion of defects~\cite{amorim2007,botello2011}, and cutting in nanoribbon samples~\cite{son2006,son2006b,pisani2007}. Another perspective is to investigate different ways to organize carbon in two-dimensional lattices different from graphene's~\cite{girao2023}. These alternative nanocarbon allotropes are commonly proposed by theory~\cite{terrones2000,Mathew2010}, but some representatives have also been successfully synthesized, such as 2D biphenylene (BPN)~\cite{Qitang2021}. Most of these systems are based on carbon $sp^2$ hybridization, where carbon atoms bind to three atomic sites. However, the study of carbon at the nanoscale has also been pushed into setups involving other hybridizations. These include, for example, systems with a mixture of $sp^2$ and $sp^3$ atoms, with pentagraphene~\cite{zhang2015}, tetrahexcarbon~\cite{ram2018}, and ph-graphene~\cite{zhang2016} as notable examples. Graphynes make up another class of systems that have been studied well before these $sp^2-sp^3$ materials~\cite{baughman1987}. Graphyne (GY) is a one-atom-thick layer where $sp$ and $sp^2$ hybridized carbon atoms coexist. They have been investigated in many structural variations through theoretical calculations, with $\alpha$-, $\beta$-, and $\gamma$-GY among the best-known examples~\cite{Kang2011,Malko2012,Wu2013}. In structural terms, GY structures are usually defined from graphene as a starting point and differ from each other by different distributions and concentrations of acetylenic units ($-C\equiv C-$), as well as by the length of these $sp$ connections ~\cite{ivanovskii2013,Wu2013,Kang2019}. The science of GYs has also gained more relevance after the synthesis of graphydiine~\cite{li2010} and a TP-GDY analog featuring a triphenylene structural unit~\cite{Matsuoka2019a}. Recent literature has witnessed the theoretical proposal of GY forms not inspired by the graphene lattice but featuring other rings such as tetragons, pentagons, heptagons, \textit{etc}.~\cite{yin2013,Nulakani2017,Wang2019,oliveira2022}. 

The proposal of new 2D $sp^2$ nanocarbon forms,  the investigation of non-graphitic GY systems, and the experimental realization of 2D byphenylene and some GY forms lead us to propose and study a GY structure based on the 2D BPN lattice. The GY we consider here is based on including a minimal acetylenic sector sandwiched between each pair of first-nearest-neighbor carbon atoms from BPN. We use density functional theory simulations to gain insight into the stability of this system and investigate its electronic structure. We demonstrate that this system preserves spin-based electronic states that were previously reported in the $sp^2$ BPN counterpart~\cite{alcon2022}. However, the spin-polarized state results in a semiconducting configuration, resulting in a system with high potential for applications in nanoelectronics and spintronics.

\section{Methods}

Our computational simulations are based on density functional theory (DFT)~\cite{hohenberg1964,kohn1965} implemented in the SIESTA code~\cite{Soler2002}. Core electrons are considered by employing Troullier-Martins pseudopotentials~\cite{troullier1991} and valence electrons are described in terms of a double-$\zeta$ polarized (DZP) basis set of numerical atomic orbitals~\cite{Soler2002}. The exchange-correlation energy is taken into account using the generalized gradient approximation (GGA) as parameterized by Perdew-Burke-Ernzerhof (PBE)~\cite{perdew1996}. Real-space integrations are performed on a grid defined by a 400 Ry mesh cutoff, and a $16\times18\times1$ Monkhorst-Pack sampling is used for reciprocal space integrations for the studied GY nanosheets. We consider $k$-sampling schemes with a similar density of $k$-points for other systems considered in this study. During optimization of the atomic structures for the calculation of electronic properties, no constraints were imposed. We considered a maximum post-relaxation force of 0.01~eV/\AA~for each atom and a maximum 0.1 GPa tolerance for stress components while optimizing the lattice parameters. We also performed Born-Oppenheimer molecular dynamics (BO-MD) simulations within SIESTA using an NVT (canonical) ensemble and a Nosé thermostat.

Phonon band structure calculations were performed with the DFT-based GPAW code~\cite{mortensen2005,enkovaara2010}. We used the finite difference method with atomic displacements of 0.01~\AA~to calculate force constants together with the PHONOPY package~\cite{togo2015}. GPAW uses the grid-based projector-augmented wave (PAW) method in the description of electron-ion interactions~\cite{blochl1994}. For manipulation and analysis of atomic simulations, GPAW uses the atomic simulation environment (ASE) package~\cite{larsen2017}. The GGA-PBE functional was also considered in the GPAW simulations. A 500~eV energy cutoff was adopted to define the plane-wave basis, and a 0.1 eV smearing was used to compute Fermi-Dirac occupations. The $3\times3\times1$ supercell considered in these simulations was sampled with $2\times2\times1$ Monkhorst–Pack points for reciprocal space integrations. The quasi-Newton method is used for atomic coordinate optimization considering a maximum threshold value of 1~meV/\AA~for atomic forces and 0.01~GPa for stress, with the structures further symmetrized with PHONOPY.

\section{Results and Discussion}

\subsection{Structural details}

\begin{figure}[ht!]
\includegraphics[width=\columnwidth]{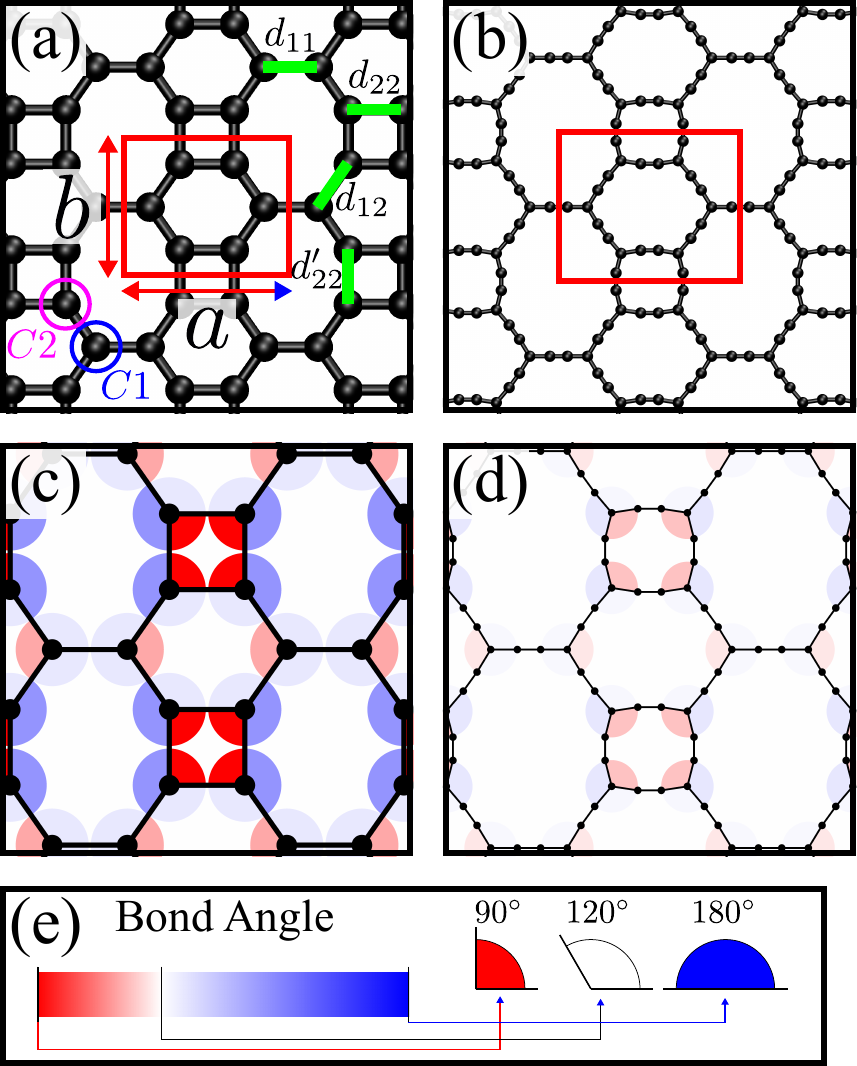}
\caption{(a) Atomic structure of the BPN monolayer, together with a representation of its primitive unit cell (red rectangle), its lattice parameters $a$ and $b$ (red double arrows), its two nonequivalent atoms, $C_1$ and $C_2$ (blue and pink circles, respectively), and the four distinct bond lengths, $d_{11}$, $d_{22}$, $d'_{22}$, and $d_{12}$ (green segments). (b) Atomic structure of the $\alpha$-BPNGY structure with its rectangular primitive cell highlighted by the red rectangle. (c) Bond angles in the BPH structure. (d) Bond angles for the $sp^2$ atoms in $\alpha$-BPNGY. (e) The color bar scale used in (c-d), with full red (blue) [white] representing the angles $90^\circ$ ($180^\circ$) [$120^\circ$].}
\label{fig-01}
\end{figure}

Fig.~\ref{fig-01}a illustrates the BPN structure, with the primitive unit cell highlighted by a red rectangle. Its lattice parameters, $a$ and $b$ (red arrows), the two non-equivalent $C1$ (blue circle) and $C2$ (pink circle) atoms in the structure, and its four different bond types (green segments) are also illustrated in Fig.~\ref{fig-01}a. The types of bonds are denoted by $d_{11}$, $d_{22}$, $d'_{22}$, and $d_{12}$, depending on the atoms involved in each bond. In that case, $d_{22}$ represents the bond length between two $C2$ atoms inside the same hexagon, while $d'_{22}$ represents a bond between two $C2$ atoms from different hexagons. These bond lengths have the following values: $d_{11}=1.451$~\AA, $d_{22}=1.461$~\AA, $d'_{22}=1.459$~\AA, and $d_{12}=1.412$~\AA. This system is organized in a rectangular 2D Bravais lattice, with vectors $\mathbf{a}_1$ and $\mathbf{a}_2$ along the $a$ and $b$ directions, respectively. These BPN lattice parameters are $a=4.53$~\AA~and $b=3.77$~\AA.

Fig.~\ref{fig-01}b illustrates the $\alpha$-GY structure based on the BPN lattice, or simply $\alpha$-BPNGY. It consists of the insertion of a minimal acetylenic chain ($-C\equiv C-$) in the middle of each and all of the original $C^{sp^2}-C^{sp^2}$ links from BPN. The $\alpha$-BPNGY monolayer has lattice parameters $a=12.78$~\AA~and $b=10.52$~\AA, with nearly the same aspect ratio ($a/b\sim1.215$) compared to BPN ($a/b\sim1.202$). For the $\alpha$-BPNGY bond lengths, we have one $C^{sp}-C^{sp}$ and two $C^{sp}-C^{sp^2}$ bonds corresponding to each $C^{sp^2}-C^{sp^2}$ connection from BPN. First, we look at the $\alpha$-BPNGY bridges analogous to the $d_{11}$, $d_{22}$, and $d'_{22}$ links in BPN. By symmetry, the two $C^{sp}-C^{sp^2}$ bonds in each of these bridges are equivalent. We find 1.411~\AA, 1.401~\AA, and 1.420~\AA~values for such $C^{sp}-C^{sp^2}$ bonds in the $d_{11}$, $d_{22}$, and $d'_{22}$ $\alpha$-BPNGY bridges, respectively. The corresponding values for the $d_{12}$ links are very similar to each other, namely 1.399~\AA~and 1.393~\AA. We note that these values are within a range between 1.393~\AA~and 1.420~\AA, which is narrower than the corresponding range of $C^{sp^2}-C^{sp^2}$ bonds in BPN (from 1.412~\AA~to 1.461~\AA). For the $C^{sp}-C^{sp}$ bonds along the $d_{11}$, $d_{22}$, $d'_{22}$, and $d_{12}$ links, we find 1.236~\AA, 1.245~\AA, 1.237~\AA, and 1.244~\AA, respectively, a range narrower than in the $C^{sp}-C^{sp^2}$ cases. This is similar to what happens in other non-graphitic GY structures~\cite{oliveira2022,vieira2024} and reflects the fact that the non-hexagonal rings are better accommodated in GY structures than in their full-$sp^2$ counterparts due to the large rings structure. All these bond lengths are listed in Table~\ref{tab1}.


\begin{table}[h!]
\caption{Bond lengths for different interatomic connections in $\alpha$-BPNGY for the NP and SP configurations.}
\centering
\begin{tabular}{|c|c|c|c|}
\hline
Bond type                     & bond length (NP) & bond length (NP)   \\
\hline                                 
$d_{11}$ ($C^{sp}-C^{sp^2}$)  & 1.411~\AA & 1.410~\AA \\
\hline
$d_{12}$ ($C^{sp}-C^{sp^2}$)  & 1.399/1.393~\AA & 1.400/1.393~\AA \\
\hline                                 
$d_{22}$ ($C^{sp}-C^{sp^2}$)  & 1.401~\AA & 1.405~\AA \\
\hline                                             
$d'_{22}$ ($C^{sp}-C^{sp^2}$) & 1.420~\AA & 1.416~\AA \\
\hline                                 
$d_{11}$ ($C^{sp}-C^{sp}$)  & 1.236~\AA & 1.236~\AA \\
\hline
$d_{12}$ ($C^{sp}-C^{sp}$)  & 1.244~\AA & 1.244~\AA \\
\hline                                 
$d_{22}$ ($C^{sp}-C^{sp}$)  & 1.245~\AA & 1.243~\AA \\
\hline                                             
$d'_{22}$ ($C^{sp}-C^{sp}$) & 1.237~\AA & 1.238~\AA \\
\hline                                             
\end{tabular}
\label{tab1}
\end{table}


The bond-angle profiles also illustrate the better accommodation of non-hexagonal rings in GY structures, especially for the 4-membered rings. In BPN, the internal angles from the tetragonal rings are $90^\circ$. This results in a local structure with strong electronic repulsion between the charge density clouds of the orthogonal $C-C$ bonds emerging from a given site, and it is likely to be the most reactive sector of the structure. In $\alpha$-BPNGY, due to a better accommodation of the interatomic bonds, the internal angles of the 4-membered cycles at the $sp^2$ sites are $\approx 113^\circ$, much closer to the $120^\circ$ angle from the corresponding ``perfect'' $sp^2$ hybridization in graphene, similar to other BPN-based GY recently proposed~\cite{vieira2024}. Overall, all the $sp^2$ atoms in $\alpha$-BPNGY have a bond angle profile closer to that of carbon atoms in graphene, compared to BPN. This is illustrated in Fig.~\ref{fig-01}c-d for BPN and $\alpha$-BPNGY, respectively, with the color scale indicated in Fig.~\ref{fig-01}e. To accommodate angles closer to the characteristic value $120^\circ$ found for $sp^2$ hybridization, the $C-C\equiv C-C$ bridges that make up the tetragonal rings are curved outward of the ring, while those bridges that link successive hexagons along the $a$ direction maintain their linear configuration.

\subsection{Stability of $\alpha$-BPNGY}

To evaluate the stability of $\alpha$-BPNGY relative to other known systems, we computed the formation energy per atom. Here, we define this energy in two different ways. First, we considered the $E^{\rm form}_1$) parameter defined as
\begin{equation}\label{eform1}
E^{\rm form}_1=\frac{E_{t}-N_{\rm atoms}\mu_C}{N_{\rm atoms}},
\end{equation}
where $E_t$ ($N_{\rm atoms}$) is the total energy of the system (number of atoms) per unit cell, and $\mu_C$ represents the energy per atom in graphene. In simple terms, $E^{\rm form}_1$ compares the energy per atom in the system with the energy per atom in graphene, used as reference. We computed $E^{\rm form}_1$ for $\alpha$-BPNGY, the conventional graphene-like $\alpha$-, $\beta$-, and $\gamma$-GYs, graphene, BPN, and a polyyne chain, with the corresponding values listed in Table~\ref{tab2} in ascending order. Taking into account the definition of Eq.~\ref{eform1}, $E^{\rm form}_1$ for graphene is set as zero as a reference, as it is the most stable system considered here. From the data in Table~\ref{tab2}, BPN is found to be the most stable system in the list after graphene. Except for a polyyne chain, $\alpha$-BPNGY is the less stable system. 

It is important to note that the computation of $E^{\rm form}_1$ considers carbon atoms $sp$ and $sp^2$ on the same footing compared to the energy per atom in graphene. One should note that the $sp$ and $sp^2$ states of carbon represent two stable hybridization states and occur spontaneously in nature. However, even though their relative stabilities are expected to be different from each other, there is a thermodynamic barrier that prevents a $sp\leftrightarrow sp^2$ transition from occurring freely. For this reason, we compared the energy of the system, taking into account its $sp/sp^2$ composition, with that of separate $sp$ and $sp^2$ reference systems. Therefore, we defined a second $E^{\rm form}_2$ quantity as
\begin{equation}~\label{eform2}
E^{\rm form}_2=\frac{E_{t}-N_{sp}\mu_{sp}-N_{sp^2}\mu_{sp^2}}{N_{\rm atoms}},
\end{equation}
where $N_{sp}$ ($N_{sp^2}$) is the number of atoms with $sp$ ($sp^2$) hybridization and $\mu_{sp}$ ($\mu_{sp^2}$) is the energy per atom in a polyyne chain (graphene). So, $E^{\rm form}$  for the $sp$ and $sp^2$ (polyyne chain and graphene) is set to zero as the reference system. Note that this gives a contrasted view of relative formation energies. For instance, graphene is now on the same comparison level as a polyyne chain (which was considered the less stable system according to $E^{\rm form}_1$). The corresponding data for $E^{\rm form}_2$ is also listed in Table~\ref{tab2} in ascending order. From this $E^{\rm form}_2$ metric, we conclude that $\alpha$-BPNGY is much closer to its pair of reference systems than BPN, as $E^{\rm form}_2$ is equal to 0.200 eV and 0.483 eV for $\alpha$-BPNGY and BPN, respectively. When compared to the traditional GY systems, $\alpha$-BPNGY is found to be slightly less stable than them, considering both $E^{\rm form}_1$ and $E^{\rm form}_2$.

\begin{table}[h!]
\caption{Formation energies per atom for $\alpha$-BPNGYs, conventional GYs, graphene, BPN, and a polyyne chain.}
\centering
\begin{tabular}{|c|c|c|c|}
\hline
System           & $E^{\rm form}_1$ (eV) & System           & $E^{\rm form}_2$ (eV)   \\
\hline                                 
Graphene         & 0.000             & Graphene         & 0.000   \\
\hline
BPN              & 0.483             & Polyyne          & 0.000   \\
\hline                                 
$\gamma$-GY      & 0.752             & $\gamma$-GY      & 0.107   \\
\hline                                             
$\beta$-GY       & 1.009             & $\beta$-GY       & 0.149   \\
\hline                                             
$\alpha$-GY      & 1.129             & $\alpha$-GY      & 0.162   \\
\hline                                             
$\alpha$-BPNGY   & 1.167             & $\alpha$-BPNGY   & 0.200   \\
\hline                                 
Polyyne          & 1.289             & BPN              & 0.483   \\
\hline                                 
\end{tabular}
\label{tab2}
\end{table}



\begin{figure}[ht!]
\centering
\includegraphics[width=\columnwidth]{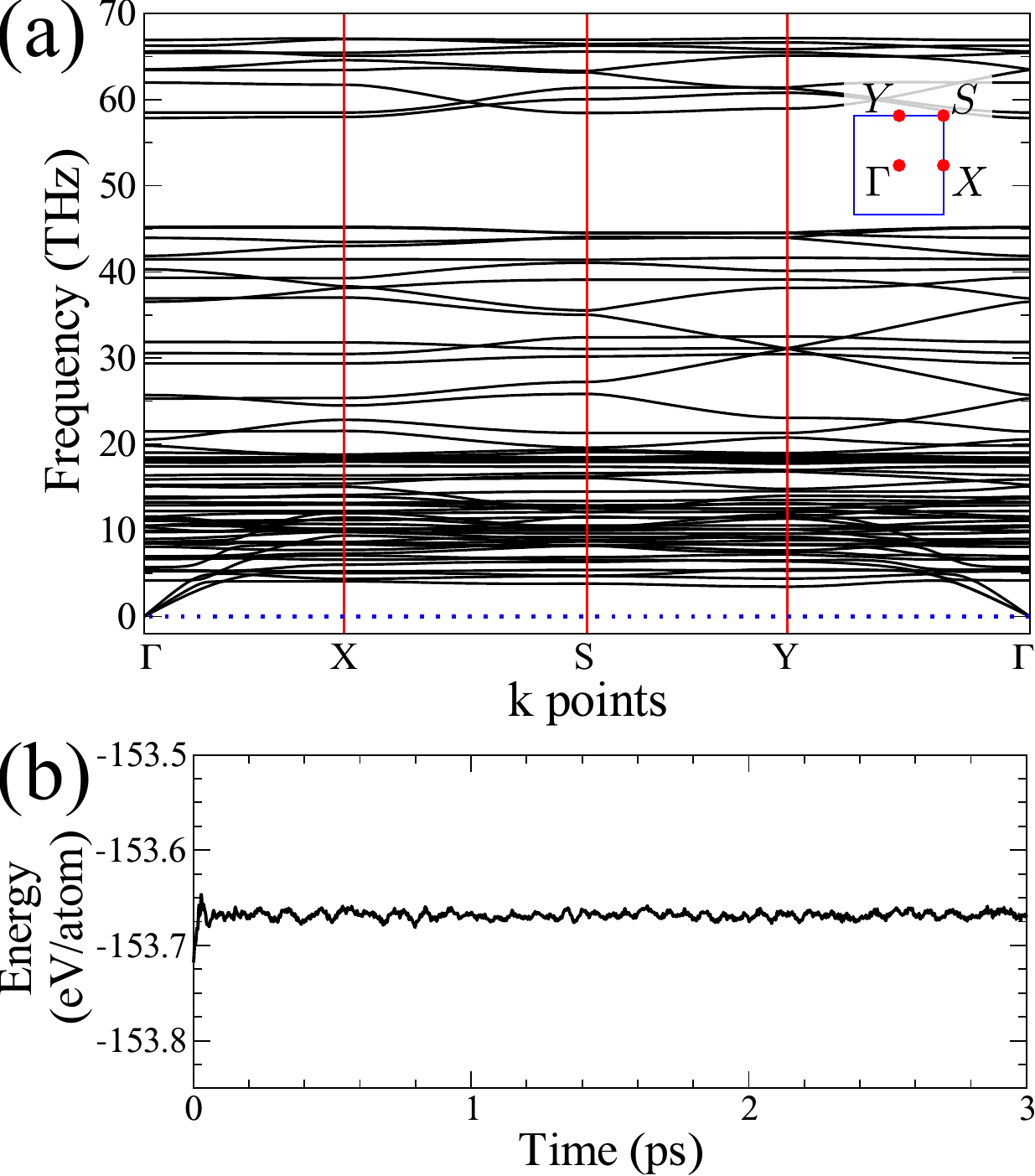}
\caption{(a) Phonon band structure for the $\alpha$-BPNGY monolayer along the high-symmetry lines of the Brillouin zone. The dynamical stability of the system is verified by the absence of complex frequencies (which would conventionally appear as negative frequencies in the plot). The acoustic branches along the $\Gamma-X$ and $\Gamma-Y$ feature a slightly different group velocity (see text). (b) Total energy, as a function of time, for $\alpha$-BPNGY according to the BO-MD simulation.}
\label{fig-02}
\end{figure}

Dynamical stability, as evaluated from the phonon band structure, is also important to consider when suggesting whether a system is likely to be synthesized. We show the phonons band structure of $\alpha$-BPNGY in Fig.~\ref{fig-02}a, where we observe no imaginary modes (which would conventionally appear as negative frequencies), indicating that the system is dynamically stable. The phonon dispersion of $\alpha$-BPNGY lacks a gap between the acoustic and optical modes. This phenomenon is commonly associated with systems that exhibit low thermal conductivity, a typical feature of materials containing acetylenic groups (graphyne-type systems), due to elevated phonon scattering within the system \cite{santos2024proposing,mortazavi2022ultrahigh,wang2016tunable}. Another characteristic commonly observed in graphyne systems is the presence of a flat dispersion band around 66~THz, which is also evident in $\alpha$-BPNGY. This phonon mode is related to the vibrations of the sp bonds in the acetylenic groups \cite{jenkins2024thd}, which are inserted between the $C_1$ and $C_2$ atoms in BPN (see Fig. \ref{fig-01}a). In addition to a flat dispersion, the range of isolated frequencies between 55 and 66~THz is directly related to the transition from BPN to $\alpha$-BPNGY, with the system showing a maximum dispersion beyond this isolated range, around 45~THz, which is close to that observed in the biphenylene network \cite{luo2021first}.

We can also assess thermal stability using BO-MD simulations. Here we considered a $3\times3$ supercell, a total simulation time of 3~ps (with a timestep of 0.5~fs), and a temperature value of 400~K. During the entire simulated dynamics, no bond breaking or formation was observed from a visual inspection of the trajectory. Beyond visual inspection of the animation, a quantitative way to evaluate the integrity of the system is the computation of the Lindemann indices: its local ($\delta_i$, $i$ ranging over the structure's atoms) and global ($\delta$) values~\cite{ding2005,singh2013}. The maximum values for $\delta_i$ and $\delta$ were less than 0.0281, which is a good indication of the integrity of the system if we compare it with other nanocarbon systems~\cite{zhang2007,silva2020,oliveira2022}. In Fig.~\ref{fig-02}b, we also show the total energy per atom as a function of time for the BO-MD simulation. Apart from characteristic fluctuations of MD simulations, we observe no abrupt changes in the energy profile, indicating the absence of structural transitions during the time evolution.

We note that neither the phonon analysis nor the molecular dynamics results take into account variations in the lattice parameters of the system. This aspect is the basis for the definition of the system's elastic tensor components: $C_{11}$, $C_{22}$, $C_{12}$, and $C_{66}$, corresponding to $1-xx$, $2-yy$, $6-xy$ according to the Voigt notation~\cite{andrew2012}. These quantities allow us to compute the elastic strain energy $U(\varepsilon)$, defined as the difference between the total energy of the strained and relaxed systems per unit area. The relation between $U(\varepsilon)$ and the $C$s components is written for low-strain values as:
\begin{equation}
U(\varepsilon)=\frac{1}{2}C_{11}\varepsilon^2_{xx}+\frac{1}{2}C_{22}\varepsilon^2_{yy}+C_{12}\varepsilon_{xx}\varepsilon_{yy}+2C_{66}\varepsilon^2_{xy}.
\end{equation}
Here, $\varepsilon_{xx}$ ($\varepsilon_{yy}$) is the strain component along the $x$ ($y$) direction, while $\varepsilon_{xy}$ represents shear stress. Considering that the lattice vectors of the relaxed system are given by
\begin{equation}
\mathbf{a}_1=(l^0_x,0)\quad\textrm{and}\quad\mathbf{a}_2=(0,l^0_y),
\end{equation}
the $\varepsilon_{xx}$ and $\varepsilon_{yy}$ strain values are defined as:
\begin{equation}
 \varepsilon_{xx}=\frac{l_x-l^0_x}{l^0_x},\quad\textrm{and}\quad \varepsilon_{yy}=\frac{l_y-l^0_y}{l^0_y},
\end{equation}
with $l^0_x$ ($l^0_y$) being the relaxed lattice constant along $x$ ($y$) and $l_x$ ($l_y$) the corresponding strained values. The $(\varepsilon_{xx}\neq0;\varepsilon_{yy}=0)$ and $(\varepsilon_{xx}=0;\varepsilon_{yy}\neq0)$ cases correspond to a uniaxial strain applied along $x$ and $y$, respectively. Biaxial strain corresponds to the $\varepsilon_{xx}=\varepsilon_{yy}$ case. The application of shear strain corresponds to the following strained vectors~\cite{major2013}
\begin{equation}
\mathbf{a}_1=(l^0_x,l^0_y\varepsilon_{xy})\quad\textrm{and}\quad\mathbf{a}_2=(l^0_x\varepsilon_{xy},l^0_y).
\end{equation}
We considered the strain values -0.01, -0.005, 0.000, 0.005 and 0.01 applied separately for uniaxial strain along $x$ and $y$, biaxial and shear strains. In the following, we used the corresponding $U(\varepsilon)$ \emph{versus} $\varepsilon$ relations to obtain the elastic tensor components through least-square fitting. The resulting values are $C_{11}=72.47$ N/m, $C_{22}=96.23$ N/m, $C_{12}=77.50$ N/m, and $C_{66}=3.04$ N/m. These values obey the Born-Huang criteria~\cite{born1954,ding2013,zhang2015}, $C_{11}C_{22}-C^2_{12}>0$ and $C_{66}>0$, predicting $\alpha$-BPNGY as a mechanically stable material. 

\subsection{Mechanical Properties}

We now compare these results regarding the elastic constants of $\alpha$-BPNGY with those of a BPN layer, for which we obtained $C_{11}=251.59$ N/m, $C_{22}=285.50$ N/m, $C_{12}=104.01$ N/m, and $C_{66}=84.77$ N/m through the same procedure applied to $\alpha$-BPNGY. In general, the elastic tensor components for the graphynic system are significantly smaller than for the full-$sp^2$ system, as $\alpha$-BPNGY has long acetylenic chains and a less dense structure, giving it a particularly higher flexible character compared to BPN. Regarding anisotropy, $\alpha$-BPNGY, and BPN are fairly similar, since $C_{22}/C_{11}\sim1.13$ for BPN and $C_{22}/C_{11}\sim1.33$ for the graphynic system. On the other hand, the $\alpha$-BPNGY $C_{66}$ value is much smaller compared to the other components of the system and to the corresponding value for BPN. It turns out that the unit cell area in shear strain varies within a much smaller range compared to uniaxial and biaxial strain. Together with the high porosity and flexibility of $\alpha$-BPNGY, this explains the very small $C_{66}$ value for the GY system.

\begin{figure}[ht!]
\centering
\includegraphics[width=\columnwidth]{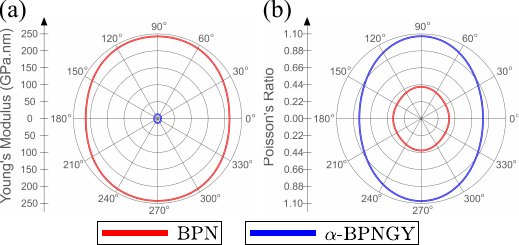}
\caption{(a) Young's modulus $Y$ \emph{versus} in-plane direction (denoted by the angle $\theta$) for BPN (red line) and $\alpha$-BPNGY (blue line). (b) Poisson's ratio $\nu$ \emph{versus} in-plane direction for BPN (red line) and $\alpha$-BPNGY (blue line).}
\label{fig-03}
\end{figure} 

In addition to the previous discussion on stability and anisotropy, the elastic tensor also allows us to determine the mechanical response of $\alpha$-BPNGY and that of the fully sp$^2$ system for comparison. From this perspective, we can obtain the Young's modulus $Y(\theta)$ and Poisson's ratio $\nu(\theta)$ through the following relations~\cite{cadelano2010}:
\begin{equation}
Y(\theta)=\frac{\Delta}{C_{11}\cos^4\theta+[\frac{\Delta}{C_{66}}-2C_{12}]\cos^2\theta\sin^2\theta+C_{22}\sin^4\theta}
\end{equation}
and
\begin{equation}
\nu(\theta)=\frac{C_{12}(\cos^4\theta+\sin^4\theta)-[C_{11}+C_{22}-\frac{\Delta}{C_{66}}\cos^2\theta\sin^2\theta]}{C_{11}\cos^4\theta+[\frac{\Delta}{C_{66}}-2C_{12}]\cos^2\theta\sin^2\theta+C_{22}\sin^4\theta},
\end{equation}
where $\Delta=C_{11}C_{22}-C^2_{12}$, and $\theta$ denotes the in-plane direction relative to the $x$ axis. 
 
Fig.~\ref{fig-03}a (Fig.~\ref{fig-03}b) plots $Y$ ($\nu$) as a function of the in-plane direction (in terms of the $\theta$ angle relative to the $x$ axis) for $\alpha$-BPNGY (blue line) and the BPN sheet (red line). For the fully $sp^2$ system, the material exhibits slight anisotropy, with $Y$ ranging from 213.7 N/m to 242.5 N/m for $\theta = 0^\circ$ (same for $\theta = 180^\circ$) and $\theta = 90^\circ$ (same for $\theta = 270^\circ$), respectively. Considering the thickness of graphene, equal to 0.335 nm \cite{shearer2016accurate}, this corresponds to a variation from 638 GPa to 724 GPa in the $x$ and $y$ directions, respectively, which are lower values than graphene's \cite{sakhaee2009elastic}. Since BPN is composed of three different carbon rings, two of which are nearly regular (hexagon and tetragon), the octagonal ring exhibits an irregular characteristic with its longitudinal direction parallel to the $y$ direction of the system (see Fig. \ref{fig-01}), making BPN stiffer in this direction. Applying tension along the $y$ direction compresses the octagonal ring further, compared to tension in the $x$ direction, which leads to the expansion of the 8-membered ring. This same geometric aspect of the system explains the Poisson's ratio of BPN, which varies from approximately $\nu = 0.36$ to $\nu = 0.41$ for $\theta = 0^\circ$ (same for $\theta = 180^\circ$) and $\theta = 90^\circ$ (same for $\theta = 270^\circ$), respectively. Similarly to the discussion on Young's modulus, the octagonal ring plays a crucial role in the expansion/compression of the nanomaterial, depending on whether the external force is applied parallel or perpendicular to the longitudinal direction of the octagonal carbon ring. Regardless of the direction of tension, $\nu > 0.25$ indicates that BPN is a ductile system, and even though $Y$ reaches about 70\% of the value reported for graphene, it differs in terms of brittleness \cite{wang2019dhq,zhang2015fracture}. It is important to note that this ductile behavior has already been reported for BPN based on fully atomistic simulations, which considered temperature effects and larger-scale systems \cite{pereira2022mechanical}.

For $\alpha$-BPNGY, the Young's modulus is generally significantly lower for the graphyne case, while the Poisson ratio is more pronounced. This result is expected for a graphyne-type system. As discussed and shown in Fig.~\ref{fig-01}, the structural degrees of freedom in graphyne-type systems tend to bring the $sp^2$ hybridizations of the nanomaterial closer to their ideal values, forming angles of $120^\circ$. When tension is applied, the energy associated with angle formation is much lower than that associated with $sp$ and $sp^2$ bonds, making the angles easily adjustable under external pressure, making $\alpha$-BPNGY a more malleable system. Consequently, this reduces the value of $Y$, which ranges from 10.1 N/m, in the $x$ direction, to 13.3 N/m, in the $y$ direction. Taking into account the graphene's thickness, these values correspond to 30.1~GPa and 39.7~GPa, respectively, which are close to the values reported for $\alpha$-GY, i.e., approximately 30.2 GPa in both directions \cite{zhang2014graphene}. On the other hand, $\nu$ varies from 0.80 to 1.07 in the $\theta = 0^\circ = 180^\circ$ and $\theta = 90^\circ = 270^\circ$ directions, respectively. Thus, $\alpha$-BPNGY is considerably more ductile than BPN, given the ease of angle rearrangement in graphyne systems, which facilitates geometric changes when tension is applied. These differences have been reported for other graphyne systems based on two-dimensional materials with $sp$ and $sp^2$ hybridization \cite{zhang2015penta,deb2020pentagraphyne,de2023nanomechanical,peng2014new,vieira2024electronic}.

\subsection{Electronic properties}

\begin{figure*}[ht!]
\includegraphics[width=\textwidth]{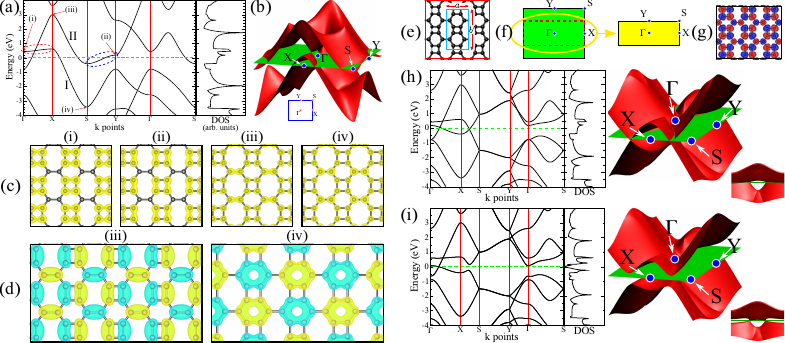}
\caption{(a) Electronic band structure of BPN along high-symmetry lines of the BZ and the corresponding DOS \emph{versus} energy profile. Bands crossing the Fermi level are identified as I and II. (b) Surface plot (red surfaces) over the entire BZ for the I and II bands highlighted in (a). The Fermi level is represented by the green rectangle. An inset also illustrates the rectangular BZ and its high-symmetry points. (c) LDOS plots for the states marked as (i), (ii), (iii), and (iv) in (a). (d) Plots for the real part of the wave-function for the (iii) and (iv) states, with the different colors representing wave-function values with opposite signs. (e) Representation of a conventional cell with two replicates of the primitive cell along the $b$ direction. (f) Representation of the original BZ (green rectangle) and the reciprocal cell (yellow rectangle) corresponding to the conventional cell represented in (e). (g) Spin-polarization plot for the SP state of BPN, with regions of spin-up (-donw) polarization represented by blue (red) color isosurfaces. (h) Same as (a-b), but for the NP state of BPN calculated on the conventional cell represented in (e). (i) Same as (h), but for the SP state of BPN.}
\label{fig-04}
\end{figure*}

In terms of electronic properties, we first revisit the electronic structure of the BPN layer so we can compare it with the properties of the  GY structure. We initially did not explicitly consider the spin degree of freedom. This spin-paired electronic configuration will be referred to as an NP (non-polarized) state and its corresponding band structure, evaluated along high-symmetry lines of the Brillouin zone (BZ), is shown in Fig.~\ref{fig-04}a together with the corresponding density of states (DOS). The plots confirm the metallic behavior of BPN previously reported in theoretical~\cite{Mathew2010,bafekry2022} and experimental~\cite{Qitang2021} studies. We observe bands crossing the Fermi level along all the high-symmetry lines, except for the $\Gamma-Y$ path. We will pay special attention to the frontier bands indicated by labels I and II in Fig.~\ref{fig-04}a. We plot these two bands over the entire Brillouin zone in Fig.~\ref{fig-04}b, with the Fermi level represented by the green rectangle. From these plots, we verify that bands I and II intercept each other by forming a pair of cone-shaped crossings along the $X-\Gamma-X$ path, at an energy $\sim0.52$~eV above the Fermi level.  Along the $\Gamma-X$ path, band I forms a low-dispersive branch (highlighted by a red dashed ellipse in Fig.~\ref{fig-04}a) associated with a narrow energy range of high DOS values. The bottom of this sector is also indicated by (i) in Fig.~\ref{fig-04}a. In the bands plot shown in Fig.~\ref{fig-04}a, we also note a slightly more dispersive branch crossing $E_F$ along the $Y-S$ path (highlighted by a blue dashed ellipse in Fig.~\ref{fig-04}a). The top of this band results in a DOS peak and is indicated by the label (ii) in Fig.~\ref{fig-04}a. We plot the local DOS (LDOS) for the energy points (i) and (ii) in Fig.~\ref{fig-04}c. These levels are strongly associated with the BPN squares. It turns out that the $90^\circ$ angles between the edges of these rings result in a region with a high electronic density due to the proximity of the electronic charges involved in these chemical bonds. The strong electronic repulsion originating from this sector is responsible for bringing electronic states close to the Fermi level of the system, as indicated by the plots in Figs.~\ref{fig-04}a-c.

Furthermore, band I (II) also has a minimum (maximum) value at the $S$ ($X$) point of the BZ. The maximum (minimum) energy value of band II (I) at the $X$ ($S$) point is also indicated by the (iii) ((iv)) label in Fig.~\ref{fig-04}a. These energy value extrema do not result in DOS peaks, but they will be relevant when compared with the BPNGY systems. For this reason, we plot their LDOS in Fig.~\ref{fig-04}c. These states are distributed over the entire structure and are not particularly limited to the squares. To gain more insight into the (iii) and (iv) states, we plot the real part of their corresponding wave-functions in Fig.~\ref{fig-04}d. The cyan/yellow colors indicate opposite signs for the wave-function amplitudes. Focusing on the hexagons in the (iii) state, we find that the wave-function phase is inverted as we move from one ring to the neighboring one along the $a$ direction. This is consistent with the position of this state in the Brillouin zone because the $X$ point is given by 
\begin{equation}
\mathbf{k}_X=\frac{1}{2}\mathbf{b}_1,
\end{equation}
and since $\mathbf{b}_1\cdot\mathbf{a}_1=2\pi$, its corresponding Bloch phase $\phi_{B}$ for the $\mathbf{R}=\mathbf{a}_1$ displacement is
\begin{equation}
\phi_{B}=e^{i\mathbf{k}_X\cdot\mathbf{a}_1}=-1.
\end{equation}
We note that the phase is the same when moving between two neighboring hexagons along the $b$ direction ($\mathbf{R}=\mathbf{a}_2$) since
\begin{equation}
\phi_{B}=e^{i\mathbf{k}_X\cdot\mathbf{a}_2}=1.
\end{equation}
In the case of the (iv) state, the wave-function sign is inverted both when moving between adjacent hexagons along the $a$ and $b$ directions, since
\begin{equation}
\mathbf{k}_X=(\mathbf{b}_1+\mathbf{b}_2)/2
\end{equation}
and 
\begin{equation}
\phi_{B}=e^{i\mathbf{k}_S\cdot\mathbf{R}}=-1,
\end{equation}
for both $\mathbf{R}=\mathbf{a}_1$ and $\mathbf{R}=\mathbf{a}_2$. 

2D BPN is composed solely of even-membered rings, namely tetragons, hexagons, and octagons. Because of this, one expects BPN to be a bipartite lattice. This means that its atoms can be split into two sets, A and B, so that all the first nearest-neighbors to any A atom come from the B set, and \textit{vice versa}. However, if we label the atoms according to the A and B sets, we end up breaking the bipartition at the edges of the primitive cell, since we necessarily find A-A and B-B bonds. To fully accommodate bipartition in the periodic BPN structure, one has to necessarily consider a conventional supercell containing 12 atoms, as represented in Fig.~\ref{fig-04}e. This is because the A/B designations in neighboring hexagons along the $b$ direction have to be swapped to obey the bipartite organization, which justifies the need for a conventional cell with two hexagons. This cell has lattice parameters given by $a=4.54$~\AA~and $b'=7.56$~\AA, with the second one being twice the value of the corresponding one in the primitive cell. The corresponding unit cell in the reciprocal space will have half the size of the original BZ along the $\Gamma-Y$ direction, as represented in Fig.~\ref{fig-04}f. This is the reason why we expect doubly-degenerated bands along the $S-Y$ path for calculations involving the conventional cell. Although the electronic properties in the NP state should not be affected in a calculation based on the conventional cell, such a setup will allow the electronic charge to converge to a different configuration in a spin-polarized calculation~\cite{alcon2022}.

Fig.~\ref{fig-04}h shows the electronic band structure of BPN in the NP state considering the doubled cell. These bands are consistent with the bands obtained in the calculation using the primitive cell (Fig.~\ref{fig-04}a). As discussed above, the bands shown in Fig.~\ref{fig-04}h are degenerated two-by-two along the new $S-Y$ path for the NP BPN sheet. In Fig.~\ref{fig-04}h we also plot the BPN's frontier bands over the entire BZ for this conventional cell setup, where we note that the $E=E_F$ states around the $S$ point in the primitive cell setup are now moved to lie in another $k$-path around the new $X$ point of the BZ. 

In the following, we show the electronic band structure of spin-polarized (SP) BPN along the high-symmetry lines of the BZ in Fig.~\ref{fig-04}i, with the corresponding DOS plot. We also show the valence and conduction bands over the entire BZ with surface plots in Fig.~\ref{fig-04}i. We first note that even though this is a spin-polarized state, the corresponding bands are spin-degenerated. This is understood in terms of the difference between the spin-up and -down components of the electronic density, represented in Fig.~\ref{fig-04}g, with blue (red) clouds representing excess spin-up (-down) charges. We note this is a configuration with zero total magnetic moment, and there is a symmetry between the spin components of the electronic charge (consistent with the lattice bipartition). This agrees with Lieb's theorem~\cite{lieb1989}, as the number of atoms is the same in the two sublattices composing the BPN structure~\cite{yazyev2010}. One notes that the spin-up and -down densities can be swapped by mirror symmetry operations along the crystallographic directions. We highlight that this SP configuration in BPN is not accompanied by any Peierls-like distortion of the atomic structure. In other words, each of the two unit cells composing the double supercell has the same bond-length and bond-angle structure as that obtained from a primitive cell-based calculation. In this sense, we conclude the use of the primitive cell leads to a magnetically frustrated (NP) configuration. On the other hand, the doubled supercell allows extra degrees of freedom, resulting in a broader set of hypothetical magnetic configurations to test. This includes the SP case discussed here and in previous literature~\cite{alcon2022}, which turns out to be the system's ground state. But here, the frustration mechanism is driven by the accommodation of the lattice bipartition only in the double supercell, rather than by lattice distortion~\cite{hou2015} or defects/temperature effects~\cite{bayaraa2021,kundu2023} as observed in other materials.

In addition, the band structure for this SP state is very similar to that of the NP case. The most remarkable change in the band structure for SP relative to the NP configuration is the opening of the bands crossing close to the $E_F$ along the $X-S$ path. This opening is followed by the other states along the $k$-curves around the $X$ point where the NP system has bands crossing the $E_F$ (as seen by comparing Fig.~\ref{fig-04}h-i). This band anti-crossing in the SP configuration collapses into the pair of cones symmetrically located along the $X-\Gamma-X$ path, with the additional effect that these cones are now moved to the Fermi level. This is also manifest in the DOS plot for the SP state, which features a zero-gap signature at $E_F$ resulting from this cone-like intersection between the valence and conduction bands of the SP BPN system. 


\begin{figure}[ht!]
\includegraphics[width=\columnwidth]{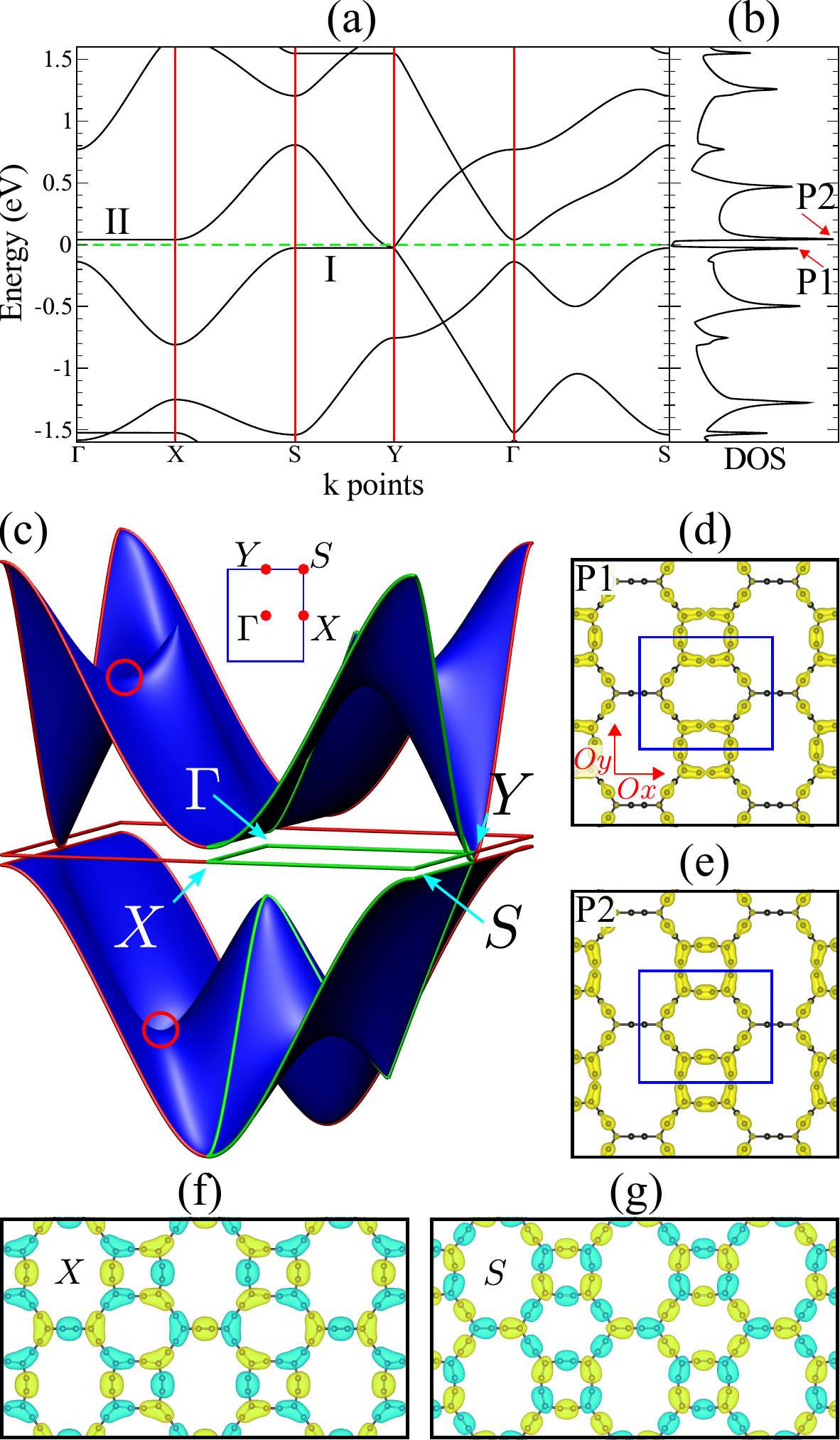}
\caption{(a) Electronic band structure of $\alpha$-BPNGY along high-symmetry lines of the BZ. The bands crossing the Fermi level are labeled I and II. (b) DOS \emph{versus} energy for the $\alpha$-BPNGY sheet. The peaks closest to the Fermi level (associated with the flat sectors of bands I and II) are indicated as P1 and P2. (c) Surface plot (blue surfaces) over the entire BZ for the I and II bands highlighted in (a). The Fermi level is represented by the red rectangle, and the high-symmetry path used in (a) is highlighted by the green lines over the Fermi level and over the I and II bands. An inset also illustrates the rectangular BZ and its high-symmetry points. (d-e) LDOS plots for the states at the P1 and P2 peaks from (b), respectively. (e-f) Plots of the real part of the wave-function for the minimum of band I at $X$, and for the maximum of band II at $S$, respectively. Different colors represent wave-function values with opposite signs.}
\label{fig-05}
\end{figure} 

Moving to the $\alpha$-BPNGY layer, Fig.~\ref{fig-05}a shows its electronic band structure, which reveals that this system is metallic. Here, the spin-degree of freedom is not considered explicitly in this NP configuration. A notable feature of $\alpha$-BPNGY's electronic structure is the presence of a pair of flat sectors from two bands close to the Fermi level ($E_F$), a feature shared with another BPNGY system~\cite{vieira2024}. These localized branches resemble the BPN less dispersive branches marked by red and blue dashed ellipses in Fig.~\ref{fig-04}a. One of these flat bands lies $\sim$ 0.04 eV above $E_F$ along the $\Gamma-X$ path, and the other $\sim$ 0.03 eV below the Fermi level along the $S-Y$ line. These bands are marked as I and II in Fig.~\ref{fig-05}a, respectively. The band II is the one that crosses the $E_F$, around the $Y$ point in $k$-space, where it has its minimum. These features can be easily seen from a surface plot of the I and II bands over the entire BZ in Fig.~\ref{fig-05}c. The flat bands result in two prominent peaks (P1 and P2) in the DOS plot of the structure, presented in Fig.~\ref{fig-05}b. The DOS also features two peaks around -0.50 and +0.46~eV, which are related to the saddle-like regions of bands I and II, respectively (as indicated by red circles in Fig.~\ref{fig-05}c). We also plot the LDOS for the two flat states of branches I and II in Figs.~\ref{fig-05}d,e, respectively. We observe that these states are distributed mainly over the atoms of the tetragonal rings, similar to the $E\sim E_F$ states in BPN. However, the electronic charge density shown in the two plots of Figs.~\ref{fig-05}d,e involve complementary sets of bonds. Namely, P1 is mostly distributed over the $C^{sp}-C^{sp}$ bonds of the $C^{sp^2}-C^{sp}-C^{sp}-C^{sp^2}$ bridges shared by tetragonal and octagonal rings, and over the $C^{sp}-C^{sp^2}$ bonds of the bridges shared by tetragonal and hexagonal rings. On the other hand, P2 extends over the $C^{sp}-C^{sp^2}$ bonds of the bridges between the tetragonal and octagonal rings, and over the $C^{sp}-C^{sp}$ bonds of the links between the tetragonal and hexagonal cycles. In addition, the flat states (P1 and P2) also spread over the other four atoms of the hexagonal rings that do not belong to the tetragonal rings. These states show a negligible amplitude over the acetylenic bonds linking successive hexagons along the $Ox$ direction. As a result, these levels look like quasi-1D states along the $Oy$ direction. This is similar to the behavior of the frontier states of naphthylene-$\gamma$ ($r4^56^410^112^1$-graphene)~\cite{beserra2020}, as well as to a BPNGY counterpart~\cite{vieira2024}. The weak participation of the $C-C\equiv C-C$ sectors linking successive hexagons along the $a$ direction compared to their counterparts in the tetragonal and hexagonal rings can be rationalized in terms of the structural details shown in Fig.~\ref{fig-01}d. The $C-C\equiv C-C$ bridges that form the tetragon and hexagon sectors are curved, bringing their corresponding levels close to the $E_F$. On the other hand, the bridges between hexagons are more stable sectors due to their perfectly linear configuration.

Another notable aspect is that the I and II bands of $\alpha$-BPNGY present a minimum at $X$ and a maximum at $S$, respectively. This is similar to the BPN case, but inverted. That is, band II has a maximum at $X$, and band I has a minimum at $S$. Such inversion behavior for the bands along the energy axis has also been reported for other graphynic systems relative to their full-$sp^2$ counterparts~\cite{oliveira2022,vieira2024}. Although these effects are not easily rationalized from the complexity of the calculation, we can examine a number of features of specific states that are consistent with the inversion. As we did for BPN, we plot the wave-functions for the minimum of I at $X$ and for the maximum of II at $S$ in Figs.~\ref{fig-05}f-g, respectively. From a chemistry point of view, these states alternate between \emph{bonding} (nonzero wave-function spreading over the bond between two atoms) and \emph{anti-bonding} (a wave-function node in the middle of the bond between two atoms) characters over the structure. The bonding character is even more explicit for the minimum of I at $X$, as it spreads over trimers (at the corners of the hexagonal rings). So, it is natural to expect that state I at $X$ has a lower energy value compared to state II at $S$. Note the wave-function of the I minimum at $X$ is odd for successive hexagonal rings along the $a$ direction, and even along the $b$ direction. This is consistent with the position of the $X$ point in the BZ and similar to the (iii) state in BPN (see Fig.~\ref{fig-04}a). In this way, the six $sp^2$ atoms of a given hexagonal ring have alternating phases for the wave-function. In addition, due to the bonding character of the wave-function lobes, the dimer between two of these atoms cannot form a \emph{bonding} pattern, otherwise, the two $sp^2$ atoms would have the same phase (opposite to the central dimer), breaking the state's symmetry. Instead, the atoms of the acetylenic dimer in the hexagon form nodal planes in their middle sector and bond lobes with the $sp^2$ sites, as shown in Fig.~\ref{fig-05}f. The stronger bonding character of state I at $X$ compared to the wave-function of band II maximum at $S$ can also be seen by the lower number of nodal planes it has. Keeping in mind the comparison with BPN, this weaker bonding (or stronger antibonding) character of band II causes it to lie above the $E_F$, while the I band moves downward, consistent with the $E$ inversion. Note that the state shown in Fig.~\ref{fig-05}g is odd for two successive hexagons both in the $a$ and $b$ directions. As the $sp^2$ atoms within a hexagonal ring have the same phase (as in BPN), the central dimers form a bonding pattern with each other and not with the $sp^2$ sites, increasing its anti-bonding character.

\begin{figure*}[ht!]
\includegraphics[width=\textwidth]{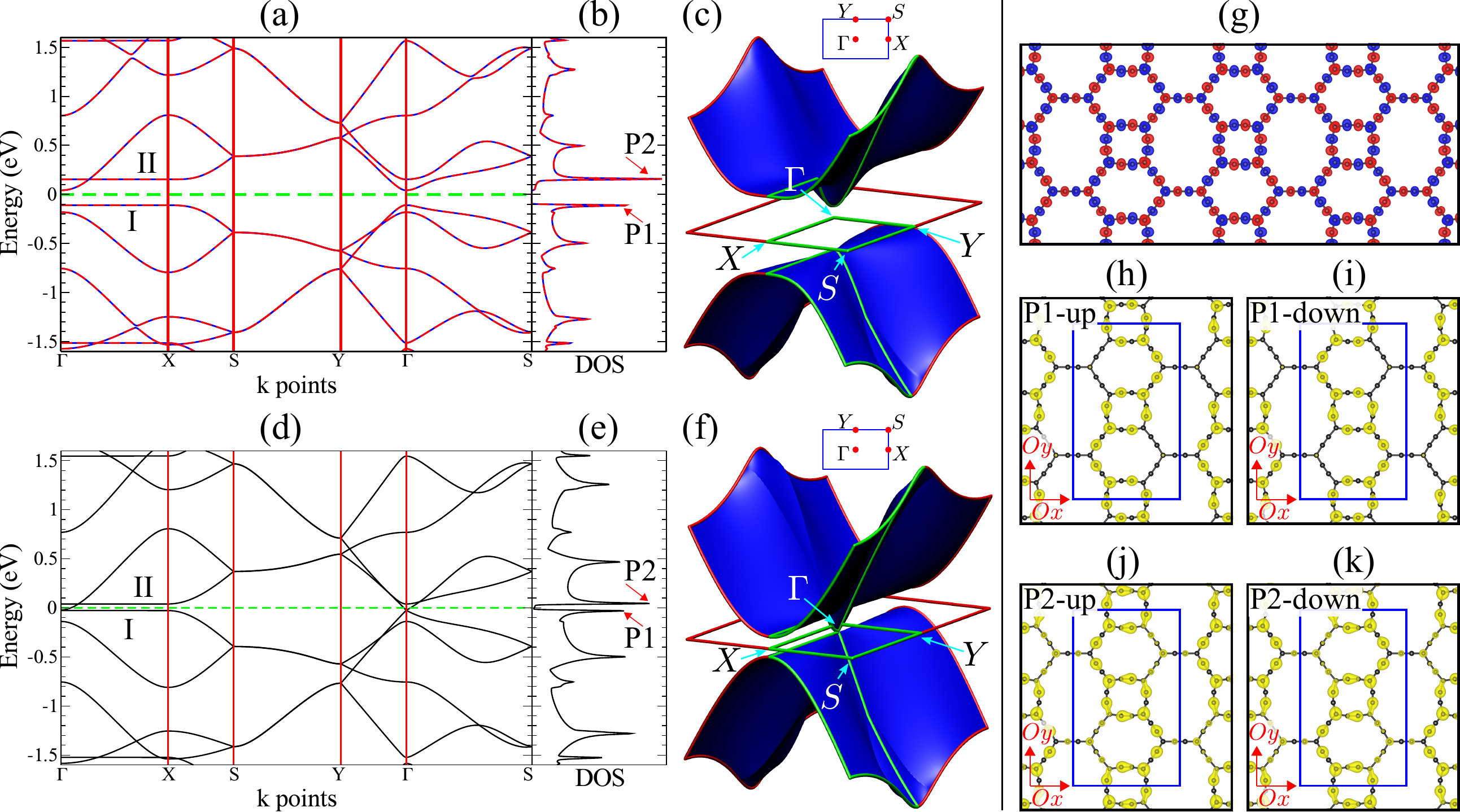}
\caption{(a) Electronic band structure of the SP state of $\alpha$-BPNGY along high-symmetry lines of the BZ. Spin-up (-down) bands are represented by full blue (red dashed) lines. This result is obtained using a supercell that corresponds to two repetitions of the primitive cell along the $b$ direction. (b) DOS \emph{versus} energy for the $\alpha$-BPNGY sheet in the SP state. (c) Surface plot (blue surfaces) over the entire BZ for the $\alpha$-BPNGY frontier bands in the SP configuration. The Fermi level is represented by the red rectangle, and the high-symmetry path used in (a) is highlighted by the green lines over the Fermi level and over the two frontier bands. An inset also illustrates the rectangular BZ and its high-symmetry points. (d-f) Same as (a-c), but for the $\alpha$-BPNGY in the NP state. The plot in (f) corresponds to the two bands crossing the Fermi energy. (g) Spin-polarization plot for the $\alpha$-BPNGY sheet in the SP state. Regions with majority spin-up (-down) are represented by blue (red) colored clouds. (h-i) LDOS plots for the spin-up and -down components of the states at the P1 peak from (e), respectively. (j-k) Same as (h-i), but for the states at the P2 peak from (e).}
\label{fig-06}
\end{figure*} 

The proximity of the P1 and P2 peaks to the Fermi level is also indicative that this system can potentially support spin-polarized states, as reported for a series of  graphitic~\cite{pisani2007,rossier2007,yazyev2010} and non-graphitic~\cite{vasconcelos2019,santos2023} nanocarbons. The spin-polarized calculations reveal a spin-unbalanced configuration for $\alpha$-BPNGY. The spin-resolved charge for this state is illustrated by the spin density plot shown in Fig.~\ref{fig-06}g. We highlight that such an SP configuration is only found in a conventional cell duplicated along the $b$ direction, similarly to the BPN case. This is related to the bipartition of the BPN lattice, as discussed previously~\cite{alcon2022}, which also applies to the $\alpha$-BPNGY structure. As in BPN, the emergence of the SP configuration is not accompanied or induced by geometrical (Peierls-like) distortions since we have the same bond length profiles for each of the two unit-cell blocks composing the doubled supercell. Regarding the details of the bond length values, they are very close to those from the NP configuration, as also listed in Table~\ref{tab1}. Here we observe that each atom featuring majority spin-up has first neighbors with majority spin-down (and \textit{vice-versa}). Despite being a state with local unbalance for the spin-resolved charges, this is an anti-ferromagnetic-type configuration with a zero total magnetic moment. The electronic band structure for this configuration is shown in Fig.~\ref{fig-06}a, together with the total DOS in Fig.~\ref{fig-06}b. The spin-up (-down) bands are represented by full blue (dashed red) lines. We note that these sets of bands are spin-degenerated, which results from the sublattice symmetry of the spin-up and -down components of the electronic charge density. To ease comparisons, we also plot the electronic band structure for the NP state computed from the conventional cell in Fig.~\ref{fig-06}d (together with the total DOS in Fig.~\ref{fig-06}e). In both cases, the flat bands close to the Fermi level (marked with I and II in the plots) lie along the $\Gamma-X$ path. However, their split is increased from 67~meV to 0.26~eV when moving from the NP to the SP case. In addition, the dispersive band that crosses band II does not cross band I, resulting in the opening of a band gap of 0.15~eV. In Fig.~\ref{fig-06}c,f we also plot the frontier bands for the SP and NP configurations, respectively, over the entire Brillouin zone. The charge redistribution resulting in the larger splitting of the I and II bands in the SP case is illustrated in Fig.~\ref{fig-06}h-k, where we plot the LDOS for the P1 and P2 peaks (from Fig.~\ref{fig-06}b) in the SP configuration. The spin-up and -down P1 states, for instance, are distributed over the same atoms as the P1 states from the NP case. However, the spin-up state (Fig.~\ref{fig-06}h) shows strong amplitude only over half of these atoms, while the spin-down state (Fig.~\ref{fig-06}i) covers the other half of these sites. In particular, the spin-up (-down) component of the P1 states lies over the atoms showing the majority spin-up (-down) in the plot shown in Fig.~\ref{fig-06}g. The situation is reversed for the unoccupied states from P2 (Figs.~\ref{fig-06}i-k). The energy difference between these two configurations is 24~meV per conventional unit cell, with the SP case as the ground state. Such a small difference is characteristic of other graphyne-like system~\cite{yue2012,oliveira2022} and suggests that a transition from one configuration to the other possibly requires going over a small energy barrier. However, a complete quantitative evaluation of this effect is beyond the scope of this work.

\begin{figure}[ht!]
\includegraphics[width=\columnwidth]{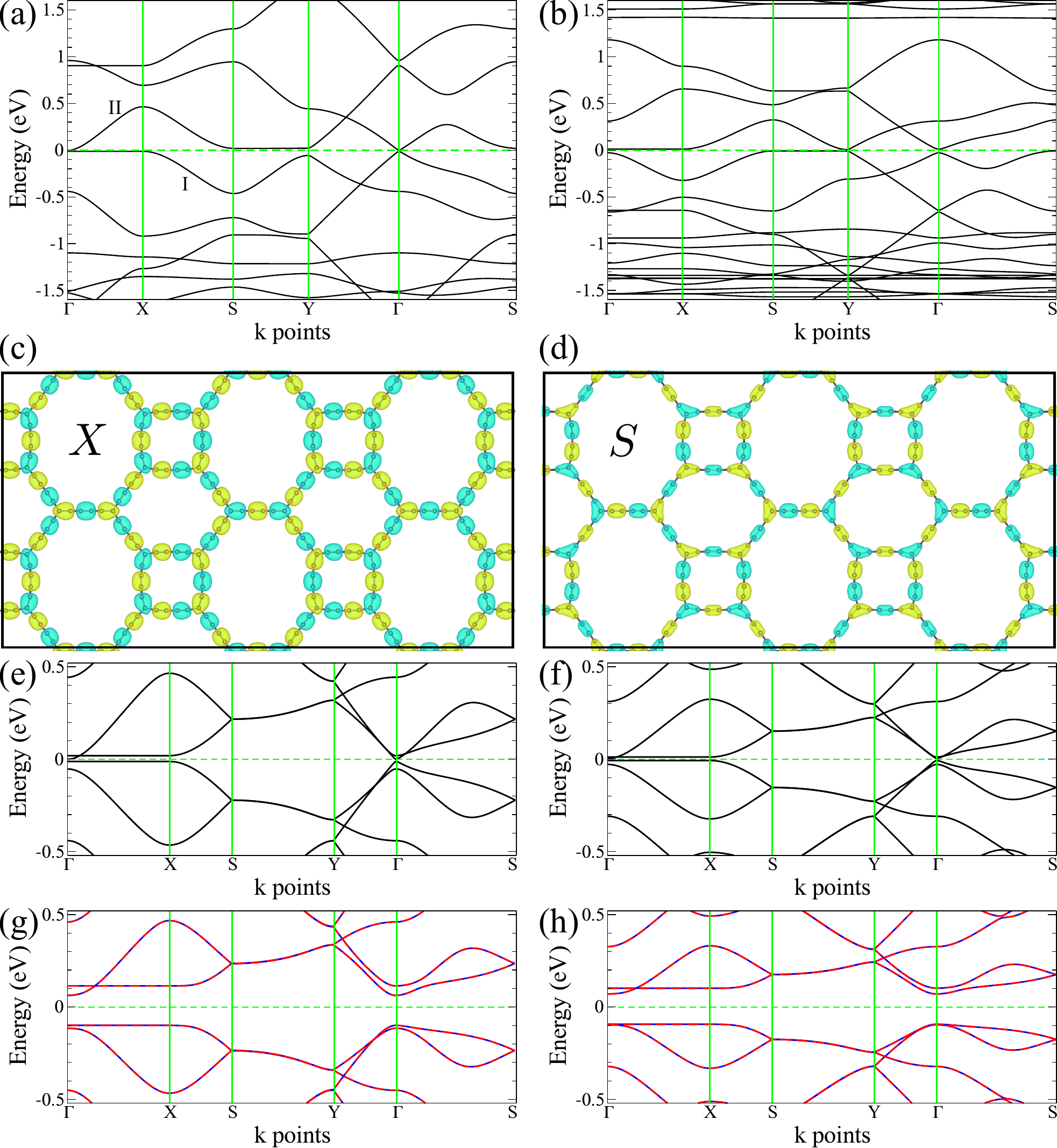}
\caption{(a) Electronic band structure of the $\alpha$-BPNGY-2 system along high-symmetry lines of the BZ. (b) Same as (a), but for the $\alpha$-BPNGY-3 case. (c-d) Plots of the real part of the wave-function for the maximum of the $\alpha$-BPNGY-2 band I at $X$, and for the minimum of the corresponding band II at $S$, respectively. Different colors represent wave-function values with opposite signs. (e) Electronic band structure of the $\alpha$-BPNGY-2 system along high-symmetry lines of the BZ in its NP configuration as obtained from a supercell consisting of two repetitions of the primitive cell along the $b$ direction. (f) Same as (e), but for the $\alpha$-BPNGY-3 case. (g) Same as (e), but for the SP configuration of the $\alpha$-BPNGY-2 sheet. (h) Same as (f), but for the SP configuration of the $\alpha$-BPNGY-3 sheet.}
\label{fig-07}
\end{figure} 

We also considered systems in which longer acetylenic chains are inserted in the system. We studied chains with two and three $-C\equiv C-$ units, which we called $\alpha$-BPNGY-2 and $\alpha$-BPNGY-3. The $\alpha$-BPNGY-2 compared to BPN is analogous to an $\alpha$-graphdiyne system compared to graphene in structural terms~\cite{niu2013,serafini2021}. Their electronic band structures in the NP state for a primitive cell configuration are shown in Figs.~\ref{fig-07}a-b. The frontier bands also feature flat bands along the $\Gamma-X$ and $S-Y$ paths. In addition, these bands show extreme values at $X$ and $S$. However, a new energy inversion pattern appears for $\alpha$-BPNGY-2 compared to $\alpha$-BPNGY, so that we have a maximum/minimum at $X$/$S$, as in BPN. For $\alpha$-BPNGY-3, a new inversion occurs and it becomes similar to the original $\alpha$-BPNGY, with a minimum/maximum at $X$/$S$. 

We can understand these differences between $\alpha$-BPNGY, $\alpha$-BPNGY-2, and $\alpha$-BPNGY-3 by examining the wave-function profile for these states at $X$ and $S$. We focus on the case of the $\alpha$-BPNGY-2 sheet, for which we show the wave-function for the maximum at $X$ and the minimum at $S$ in Fig.~\ref{fig-07}c and d, respectively. Looking back to the state at $X$ for $\alpha$-BPNGY (Fig.~\ref{fig-05}f), we observe wave-function lobes over the $C^{sp}-C^{sp^2}-C^{sp}$ trimers at the vertices of the hexagonal rings, as well as subsequent lobes around these sectors that have opposite signs for the wave-function (with nodal regions between them). To construct $\alpha$-BPNGY-2 from the $\alpha$-BPNGY structure, we insert additional $C\equiv C$ sectors (between the $C^{sp}-C^{sp^2}-C^{sp}$ trimers). This introduces a \emph{bonding}-like lobe over such a central dimer. As this dimer lobe in $\alpha$-BPNGY-2 is limited by two nodal planes on both sides of the chain, its wave-function value has an opposite sign relative to their neighboring trimer lobes at the vertices of the hexagonal rings. As a consequence, all the trimer lobes inside a hexagon now have the same sign. This is exactly the profile for the wave-function of the band I minimum at the $S$ point of the BZ for $\alpha$-BPNGY-2. In other words, the symmetry of the bonding state at $X$ for $\alpha$-BPNGY (Fig.~\ref{fig-05}e) now becomes compatible with the symmetry of the $S$ point in $\alpha$-BPNGY-2 (Fig.~\ref{fig-07}d). A similar argument explains the symmetry of the state at $S$ in $\alpha$-BPNGY. As a result, it becomes compatible with the symmetry of the $X$ point in $\alpha$-BPNGY-2 as the wave-function lobes at the vertices of the hexagons now have opposite signs (Fig.~\ref{fig-05}c). A similar comparison between $\alpha$-BPNGY-2 and $\alpha$-BPNGY-3 explains why the latter system has a band profile qualitatively similar to $\alpha$-BPNGY. Such an inversion mechanism is relevant for the $\alpha$-BPNGY structures, as their bands have a narrower dispersion compared to BPN and these states become closer to the $E_F$.


These systems also feature a semiconducting SP state within the conventional cell (doubled along the $b$ direction), showing band gaps of nearly 0.16~eV for both $\alpha$-BPNGY-2 and -3. These bands are shown in Figs.~\ref{fig-07}g-h. For comparison purposes, we also plot the bands in the NP configurations in the conventional cell setups in Fig.~\ref{fig-07}e-f. The SP state is also the lowest energy configuration for $\alpha$-BPNGY-2 and $\alpha$-BPNGY-3. Due to the longer acetylenic chains in these structures, the energy lowering in SP compared to the NP states increases to 36~meV and 44~meV in the $\alpha$-BPNGY-2 and $\alpha$-BPNGY-3 cases, respectively.

\section{Conclusions}
We studied a graphyne form with the maximal number of minimal acetylenic chains possibly inserted into a biphenylene lattice. We show that this system is metallic with two highly localized states close to the Fermi level for a non spin-polarized calculation. These localized states are primarily distributed over a specific crystallographic direction of the atomic layer, while overlap is negligible between successive electronic charge densities along the second (orthogonal) direction of the lattice. This is directly related to a local bending structure shown by a subset of the system's acetylenic bridges, which do not show a perfect linear configuration along the direction of larger dispersion. Further, we demonstrate that this structure hosts a more stable spin-polarized configuration, where the frontier bands split to open a band gap. The profile of the charge distribution in such an electronic configuration involves a conventional cell twice as long as the primitive cell along the dispersive direction. This is an anti-ferromagnetic-type spin configuration reminiscent of a similar spin-polarized state found in the biphenylene full-$sp^2$ counterpart. We also show that the symmetry of some frontier states can be altered by increasing the size of the acetylenic links. This is due to the accommodation of the corresponding wave-function lobes, which change the overall symmetry of the levels as we introduce more $-C\equiv C-$ units.

\section{Acknowledgements}
P.V.S. acknowledges support of CNPq (Process No. 312548/2023-0). E.C.G acknowledges the support of the Brazilian agency CNPq (Process No. 309832/2023-3). E.C.G. and P.V.S. acknowledge support from Funda\c c\~ao de Amparo \`a Pesquisa do Estado do Piau\'i (FAPEPI) and CNPq through the PRONEM program. M.L.P.J. acknowledges the financial support of the FAP-DF grant 00193-00001807/2023-16. The authors thank the Laborat\'orio de Simula\c c\~ao Computacional Caju\'ina (LSCC) at Universidade Federal do Piau\'i for computational support. The authors also acknowledge computational support from Centro Nacional de Processamento de Alto Desempenho at Ceará (CENAPAD-UFC) and São Paulo (CENAPAD-SP).

\end{document}